\pdfoutput=1
\documentclass[aps,prl,preprint,superscriptaddress]{revtex4}
\usepackage{graphicx}% Include figure files
\usepackage{dcolumn}% Align table columns on decimal point
\usepackage{bm}% bold math
\usepackage{color}
\usepackage[thinspace,thinqspace,amssymb,pstricks]{SIunits}

\newcommand{\fig}[1]{Fig.~\ref{#1}}
\newcommand{\eq}[1]{Eq.~(\ref{#1})}

\usepackage{color}

\begin{document}

\title{Direct observation of Space Charge Dynamics by picosecond Low Energy Electron Scattering}

\author{C. Cirelli}
\affiliation{Physik-Institut, Universit\"{a}t Zurich-Irchel, Winterthurerstrasse 190, CH-8057 Z\"{u}rich, Switzerland}
\affiliation{Physikalisch-Chemisches Institut, Justus Liebig Universit\"{a}t, Heinrich-Buff-Ring 58, D-35392 Giessen, Germany}
\author{M. Hengsberger}
\affiliation{Physik-Institut, Universit\"{a}t Zurich-Irchel, Winterthurerstrasse 190, CH-8057 Z\"{u}rich, Switzerland}
\author{A. Dolocan}
\affiliation{Physik-Institut, Universit\"{a}t Zurich-Irchel, Winterthurerstrasse 190, CH-8057 Z\"{u}rich, Switzerland}

\author{H. Over}
\affiliation{Physikalisch-Chemisches Institut, Justus Liebig Universit\"{a}t, Heinrich-Buff-Ring 58, D-35392 Giessen, Germany}
\author{J. Osterwalder}
\affiliation{Physik-Institut, Universit\"{a}t Zurich-Irchel, Winterthurerstrasse 190, CH-8057 Z\"{u}rich, Switzerland}
\author{T. Greber}
\affiliation{Physik-Institut, Universit\"{a}t Zurich-Irchel, Winterthurerstrasse 190, CH-8057 Z\"{u}rich, Switzerland}

%\date{\today}

\begin{abstract}

The electric field governing the dynamics of space charge produced by high intensity femtosecond laser pulses focused on a copper surface is investigated by time-resolved low-energy-electron-scattering. The pump-probe experiment has a measured temporal resolution of better than $\unit{35}{\pico\second}$ at $\unit{55}{\electronvolt}$ probe electron energy. The probe electron acceleration due to space charge is reproduced within a 3-dimensional non-relativistic model, which determines an effective number of electrons in the space charge electron cloud and its initial diameter. Comparison of the simulations with the experiments indicates a Coulomb explosion, which is consistent with transients in the order of $\unit{1}{\nano\second}$, the terminal kinetic energy of the cloud and the thermoemission currents predicted by the Richardson-Dushman formula.
\end{abstract}

\pacs{xxxx}

\maketitle

%\section{INTRODUCTION}

The current which can be drawn from a hot cathode follows the Richardson-Dushman formula, where the current density is given by the filament temperature and the work function \cite{Rich-Dush}.
In such a quasi-static situation the electric fields due to the space charge of the emitted electrons decrease the thermoemission \cite{Child-Langmuir}.
In pulsed electron sources, as e.g. needed for free electron lasers \cite{FEL,FEL2} or field emitters \cite{Hommelhoff06}, laser pulses may be used to heat and emit electrons. 
In this case the field of the space charge has a transient behaviour and will, depending on its timing, also accelerate electrons.
%It is therefore of importance to understand the space charge dynamics.
There are a wealth of laser experiments that address the heating of an electron gas with laser pulses \cite{Schoenlein87,Fann92,Audebert94,Hertel96} and the concomitant space charge problem \cite{passlack06}. 
In these experiments the system is excited with an optical pump pulse and its evolution is probed after a given delay with another optical pulse, where emitted light \cite{Audebert94} or electrons are used to probe the electron dynamics inside and outside of the solid. Although recent experiments with high spatial and temporal resolution show a relaxation of the space charge limitations due to strong extraction fields and localisation of the emission to hot spots \cite{Kubo07} the investigation of space charge is pivotal for high flux and energy resolution electron probes. So far there are no experiments that report transient measurements of the electric fields due to emitted electrons.\\
Here we report electron scattering experiments where ps pulses of $\unit{55}{\electronvolt}$ electrons were used to probe the dynamics of the emitted space charge outside the solid. 
It is a significant extension of the family of light-electron pump-probe experiments, which have, so far only been conducted with keV probe electrons \cite{Williamson84,Elsayed90,Ihee01,Siwick03,Krenzer06}.
If the problem of lower time resolution and fluence is solved, the use of low energy electrons has obvious advantages due to higher electron scattering cross sections and a better absolute electron energy resolution, which then allows to time resolve the energy of backscattered electrons on the meV scale.
\begin{figure}[tbh]
	\begin{center}
		\includegraphics[width=0.9\textwidth]{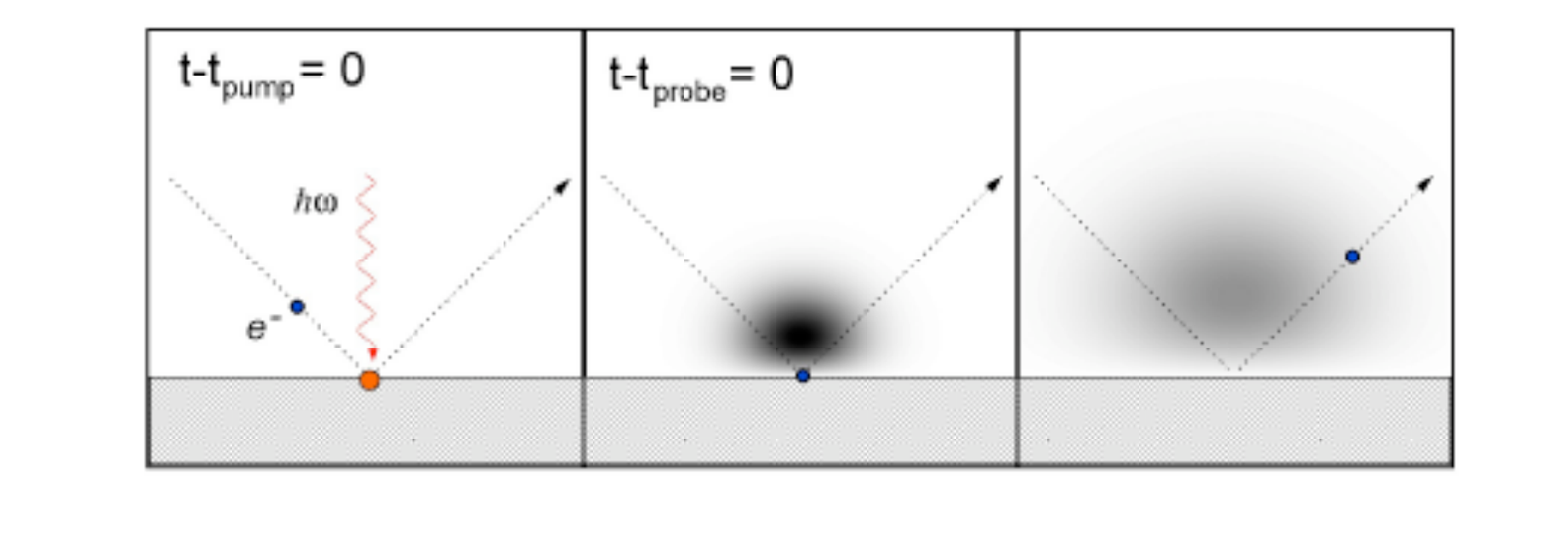}
\caption{Schematic view of the space charge experiment for 3 time frames. At $t-t_{pump}=0$ a laser pulse $\hbar\omega$ hits the surface. At $t-t_{probe}=0$ the probe electron hits the surface. The shown probe electron $e^-$ has a positive delay  $t_D=t_{probe}-t_{pump}>0$. The pump pulse excites the electrons in the solid, where a small portion of the hot electron gas may escape into the vacuum and expand. The field of the electron cloud accelerates the probing electrons while the magnitude of the effect  depends on the number of electrons in the cloud and the pump-probe delay.}
		\label{F1}
	\end{center}
\end{figure}

After improving the design of our pulsed low energy electron gun \cite{Karrer01} in terms of fluence by 3 orders of magnitude the evolution of space charge could be studied in recording the energy gain of $\unit{55}{\electronvolt}$ electrons specularly scattered off a Cu(111) surface.
\fig{F1} shows the scheme of the experiment:  space charge is created by an intense laser pulse on a surface, where the evolving electron cloud is probed with a pulsed electron beam.
At laser intensities in the GW/cm$^2$ range in a first stage the emitted charge is self-accelerated due to a Coulomb explosion \cite{Petite92} and then expands with superthermal energies into the hemisphere above the surface. 
Our experiment complements and confirms this picture with time resolved measurements of the transient electrostatic potential of a space charge cloud.

%Reference of energy broadening due to space charge (with also Montecarlo simulations) is \cite{Boemmels01}.\\
%A review of the Richardson-Dushman equation might be in \cite{Rich-Dush}, but still need to get the paper and check.\\
%Paper about "determination of space charge using random emitting conditions" \cite{Iiyoshi06} (sounds nice!), actually I didn't read it, it is in the folder.\\
%Paper from Riffe et al.\cite{Riffe92}.\\

%\section{EXPERIMENTAL SETUP}

The experiments were carried out in an Ultra-High-Vacuum (UHV) system coupled to a femtosecond laser system. All measurements were performed at room temperature at a base pressure in the $\unit{10^{-11}}{\milli\bbar}$ range. The laser system consists of a commercial Coherent MIRA Ti:sapphire oscillator, which emits pulses centered around the wavelength $\lambda_{o}=\unit{800}{\nano\meter}$ with a spectral width $\Delta \lambda \sim \unit{28}{\nano\meter}$ and a time-width $\Delta t$ of about $\unit{55}{\femto\second}$. The output pulses can be amplified to higher pulse energies by a chirped-pulse Regenerative Amplifier (Coherent RegA 9050); after the amplification process, the pulse energy is $\sim \unit{5}{\micro\joule}$/pulse at a repetition rate of $\unit{250}{\kilo\hertz}$.
The $\unit{800}{\nano\meter}$ laser light is split by a beamsplitter into two beams, one of which is directly sent towards the vacuum chamber, while the other is frequency doubled. The $\unit{400}{\nano\meter}$ beam passes through a delay stage, which can vary its optical path up to $\unit{60}{\centi\meter}$ and is then used to produce the electron pulses on a back-illuminated gold cathode of a home-built electron gun.
%
%The electron gun photocathode consists of a thin atomically flat gold film ($\unit{20}{\nano\metre}$ thick) grown on a $\unit{400}{\micro\metre}$ thick sapphire substrate. The electrons produced by the laser pulses are accelerated over a short distance (only $\unit{50}{\micro\metre}$) towards a grounded gold mesh ($\unit{5}{\micro\metre}$ thick, 44\% geometrical transmission), acting as anode. An additional $\unit{300}{\micro\metre}$ thick electrostatic lens is placed after the anode in order to focus the electrons towards the sample. The typical distance cathode-sample is in the order of few millimeters and a shield electrode ensures a field-free region for the electrons outside the head of the gun.\\
Separate experiments show that an electron yield of $\approx$1 electron/pulse at a pulse energy of $\unit{1}{\nano\joule}$ and a time resolution  better than $\unit{35}{\pico\second}$ at $\unit{55}{\electronvolt}$ primary energy can be achieved if the laser power is held below the threshold of $\unit{1}{\nano\joule}$ in order to avoid electron pulse spreading due to space-charge effects in the electron gun \cite{Cirelli07}.\\
The space-charge dynamics above a Cu(111) surface were studied in an experimental setup where the specular electron beam is back-scattered and detected (see \fig{F1}). 
The probe electron energy of 55 eV was optimized for maximum reflectivity of the (0,0) spot and low spectral overlap with the cloud electrons, where an energy of $\unit{55}{\electronvolt}$ was chosen for the experiments.
In principle any scattering geometry could be chosen, but at this energy the specular beam is at least 5 times more intense than any other diffracted beam. The scattered electrons were energy analyzed with a  $\unit{100}{\milli\meter}$ mean radius hemispherical electron analyzer (Clam2).
The pump beam is chopped with a frequency of about $\unit{20}{\hertz}$, and for each delay stage position the data are recorded with and without the pump beam, which allows to check the stability of the electron gun. 
%The ``point-by-point" normalization procedure permits to remove effects due to laser instabilities.\\
The temporal coincidence (delay zero) and spatial overlap between the light pump and the electron probe is determined with a reference curve measured with the electron-photon correlator presented in Ref. \cite{andrei}.\\
%\section{RESULTS}

\begin{figure}[tbh]
	\begin{center}
		\includegraphics[width=0.7\textwidth]{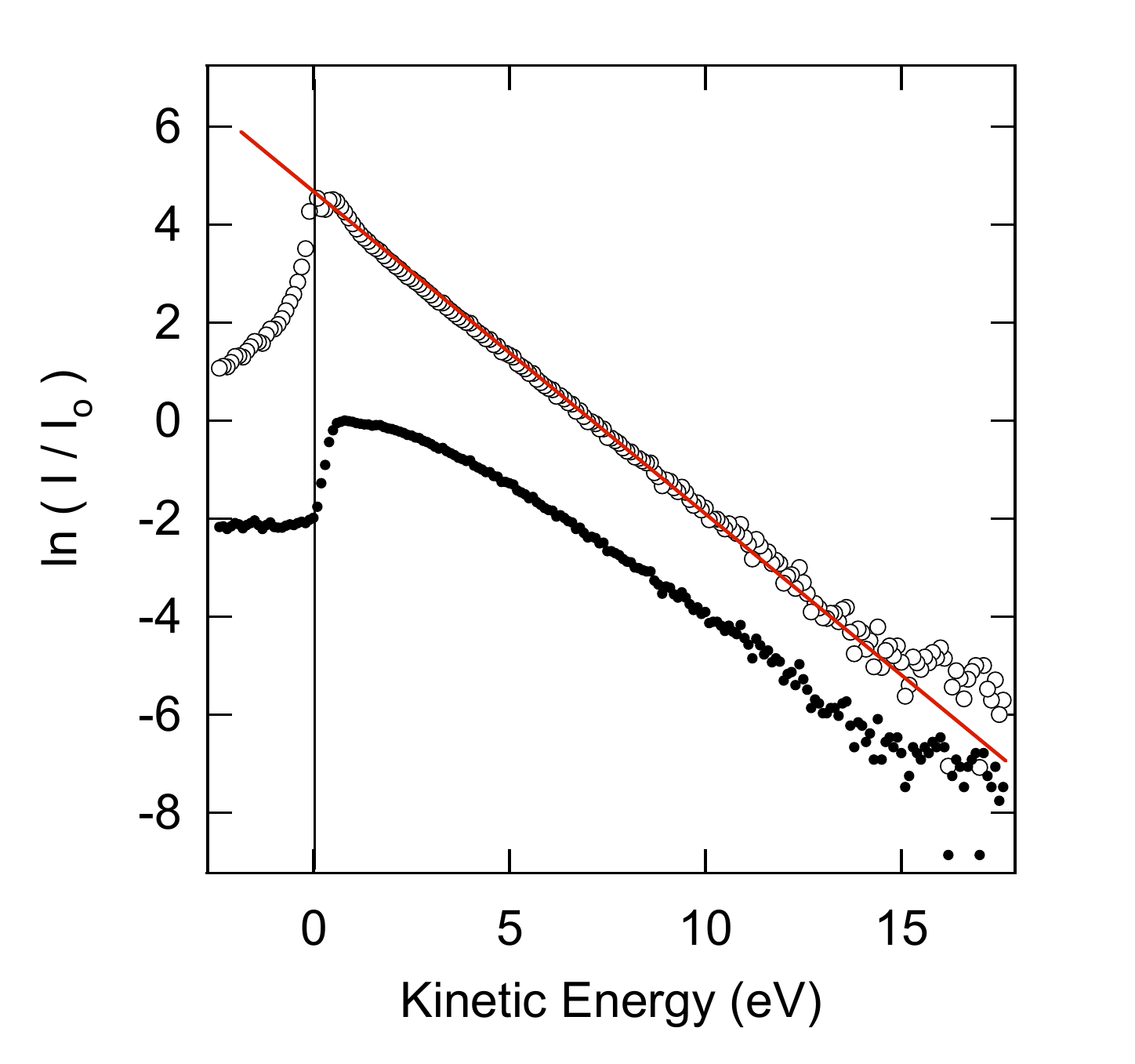}
		\caption{Logarithmic spectrum of electrons emitted by 1.5 eV,  5 $\mu J $, 100 fs laser pulses hitting a Cu(111) surface. If the spectrum (black dots) is normalized with $\sqrt{E_{kin}}$, a Boltzmann-type distribution with an effective energy k$_B T_c$ of $\unit{1.6}{\electronvolt}$ is found (open circles). $I_m$ is the maximum intensity in the measured spectrum.}
		\label{F2}
	\end{center}
\end{figure}
%
%The main drawback is the loss of the information carried by the electrons that undergo a higher-order diffraction process, but most likely the space-charge induced effects are the same for electrons of any order of diffraction, except for the different scattering angles.\\
%For the purpose of studying the specular reflection the polar angle was set to $\unit{50}{\degree}$ with respect to the sample normal, as deduced by the polar scan shown in \fig{cu111_10spot}a. An additional $\unit{0.5}{\milli\meter}$ diameter pinhole was drilled into the aluminum plate $\unit{45}{\degree}$ off-normal: in this way the electron beam can pass trough the hole also when the manipulator is tilted in order to allow the specular beam to be detected.\\
First, the electron energy distribution of the space charge that leaves the sample was measured.
In \fig{F2} the electron energy spectrum $I(E)$  of  a space charge cloud from Cu(111) for focused laser pulses of  5$\mu J$ is shown on a logarithmic scale.
The spectrum has no signature of multi photon excitations, which would be reflected in distinct steps separated by the principal photon energy of 1.5 eV.
Instead, a smooth distribution that peaks near the secondary cut-off is seen. 
If the measured electron distribution is normalized with the escape probability of the electrons, we find an almost perfect exponential distribution.
The escape probability is given by electron refraction at the inner potential and is proportional to the momentum of the electrons in the vacuum, i.e. the square root of their kinetic energy \cite{Rich-Dush,Hagstrum54,SSR_Greber}. The slope of the $\ln({I/\sqrt{E_{kin}})}$ vs. $E_{kin}$ curve translates in an average energy $k_BT_c$ of the cloud electrons of 1.6 eV. In a Boltzmann picture this energy correspoonds to  a "cloud temperature" $T_c$ of  $2\cdot10^4ÊK$. 
With this temperature a sample current may be calculated from the Richardson-Dushman equation. The current density $j$ of thermoemitted electrons gets
\begin{equation}
\label{E1}
j=AT^2\exp(-\Phi/k_BT)
\end {equation}
with $ A=1.2\cdot10^{6}$ A/K$^2$m$^2$.
It would predict for a thermal energy of $k_BT_c$=1.6 eV, a work function $\Phi$ of $\unit{4.94}{\electronvolt}$, a focus size of $\unit{0.05}{\milli\meter\squared}$, a pulse duration of $\unit{100}{\femto\second}$ and a repetition rate of $\unit{250}{\kilo\hertz}$, a sample current of about $\unit{25}{\milli\ampere}$, which is way beyond the measured currents.

The large kinetic energy in the cloud  is also at some contrast to measured electron temperatures in the solid after femtosecond laser heating which is in the order of a few $\unit{10^3}{\kelvin}$ \cite{Riffe92}. 
It is, thus,  an indication that the emitted electrons undergo a kind of self-acceleration in the course of their way to the detector.
The energy for the self-acceleration stems from the Coulomb energy in the cloud.
This picture of a Coulomb explosion \cite{Petite92} is substantiated by the measurement of acceleration of probing electrons, as will be shown in the following.

\begin{figure}[t]
	\begin{center}
		\includegraphics[width=0.9\textwidth]{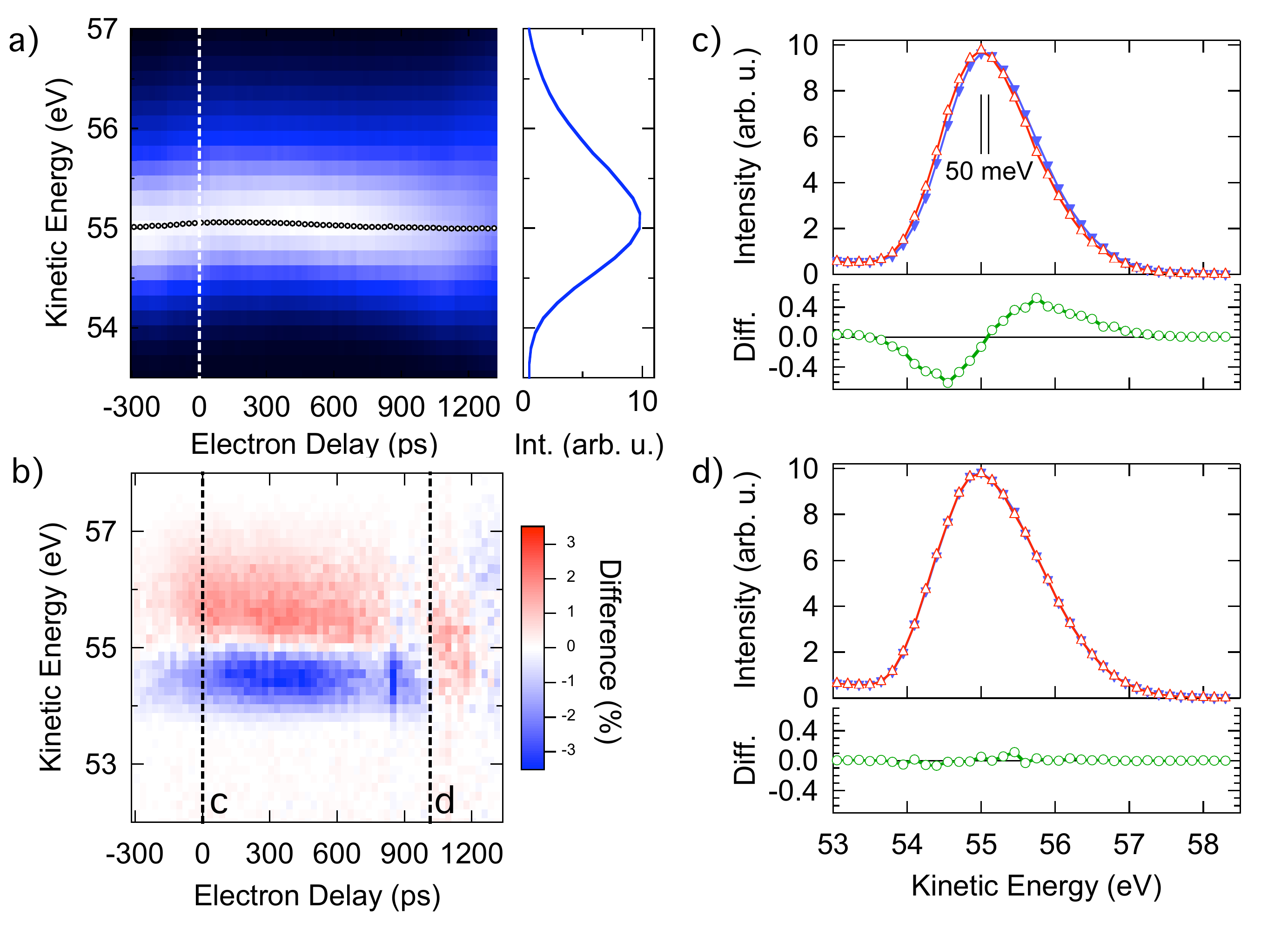}
		\caption{(Color online) Spectra of specularly scattered electrons with $\unit{55}{\electronvolt}$ kinetic energy off Cu(111)  in the "presence" of pump laser excitation with a pulse energy of $\unit{5}{\micro\joule}$. a) Specular beam intensity with pump beam as a function of delay $t_{probe}-t_{pump}$ between pump and probe pulse. The black open circles follow the peak maximum, where a time-dependent shift is noticed. On the right-hand-side the energy spectrum sliced out at the 0-delay (marked with the dashed line) is shown. b) Difference of the electron energy distribution $I_w-I_0$, where $I_w$ is the intensity of the scattered electrons with pump and $I_0$ that without pump beam. c) Energy spectra with (blue) and without (red) pump at coincidence (dashed line in b labeled 'c'). Note the shift of about $\unit{50}{\milli\electronvolt}$ in the presence of the pump pulse. The bottom panel shows the difference between the two curves. d) Energy spectra with (blue) and without (red) pump about $\unit{1}{\nano\second}$ off coincidence (dashed line in b, labeled 'd'). }
		\label{F3}
	\end{center}
\end{figure} 
\fig{F3} shows the results of the pump-probe experiment from the  Cu(111) target. Pump pulses with a duration of  $\unit{100}{\femto\second}$, a wavelength of $\unit{800}{\nano\meter}$ and a pulse energy of $\unit{5}{\micro\joule}$ and probe pulses with $\unit{55}{\electronvolt}$ electrons were used. 
\fig{F3}a) displays the electron  beam intensity as a function of the delay between light-pump and electron-probe pulses. In order to highlight transient changes, the difference $I_w-I_0$  is shown in \fig{F3}b) as a function of the delay between pump and probe, where negative delays mean that the electrons are hitting the sample surface before the pump pulses. $I_w$ is the intensity of the elastically scattered electrons when the pump light is on the sample and $I_0$ the one without pump light.
The energy spectrum of the probe electrons is affected by the pump pulse: we observe a delay-dependent energy shift towards higher kinetic energies.
In \fig{F3}c), at delay zero, this shift amounts to $\unit{50}{\milli\electronvolt}$, while after $\unit{1}{\nano\second}$ delay the probing electrons are not accelerated anymore (\fig{F3}d).
This allows to estimate the order of magnitude of the average accelerating field.
From the duration of the transient of $\unit{1}{\nano\second}$ and the energy gain of $\unit{55}{\electronvolt}$ electrons we find an average accelerating field of about $\unit{10}{\volt\per\meter}$, which corresponds to a surface charge density in a plate capacitor %$\sigma=\epsilon_oE_\perp$ 
of about 10$^9$e/m$^2$.

\begin{figure}[t]
	\begin{center}
		\includegraphics[width=0.7\textwidth]{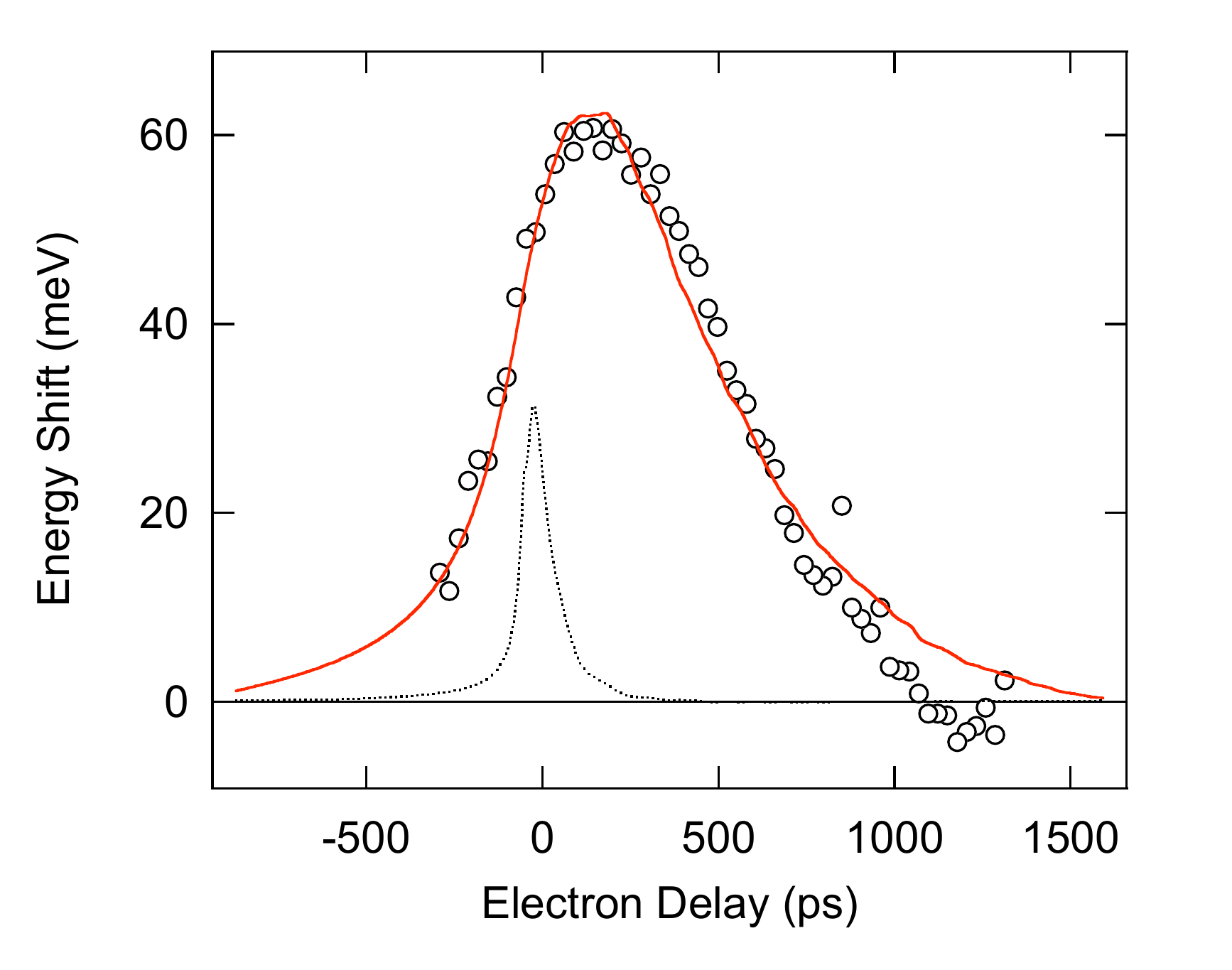}
		\caption{Kinetic energy gain of probe electrons ($E_{kin}$=55 eV) in the presence of a space charge cloud as a function of electron delay. Positive delays mean that the pump pulse hits the surface before the probe. A maximum shift of $\unit{60}{\milli\electronvolt}$ occurs about $\unit{100}{\pico\second}$ after coincidence.
The red solid line is the best fit to the model with an energy distribution as shown in \fig{F2}, with $6.3\cdot10^{4}$ electrons and an initial diameter of 1.8 mm.
The dotted line is the result for 5000 electrons in a cloud with 0.4 mm initial diameter. }
		\label{F4}
	\end{center}
\end{figure}

In order to get a more quantitative picture, the energy shifts of the probing electrons shown in \fig{F3}a) were determined for all delays by fitting Gaussians to the energy distribution curves. \fig{F4} shows the corresponding peak shift of the electron spectra with pump light relative to that without pump light. 
The data peak after delay zero and the transient is non Gaussian, i.e. has a slower decay compared to the rise at negative delays. 
It has to be noted that in contrast to light probes the rise at negative delays, i.e. when the probe electrons hit the surface before the pump pulse,  is not due to the time resolution of the experiment. 
It merely shows that the field of the expanding space charge behind, also reaches the probe electrons. 

The observed probe electron acceleration due to space charge is simulated by a non-relativistic model. This is justified since all relevant electron velocities were much smaller than the speed of light.
The  model is 3-dimensional but neglects acceleration of the cloud.  
The cloud is simulated as a homogeneously charged disk of electrons which starts to expand from the surface at a time $t-t_{pump}$=0. 
The electron energy distribution is proportional to ${\sqrt {E_{kin}} }\exp{(-E_{kin}/k_BT_c)}$, where the energy $k_BT_c$=1.6 eV is taken from the measured kinetic energy distribution in \fig{F2}. The angular distribution of the electrons is assumed to be proportional to $\cos{(\theta)}$, where $\theta$ is the polar emission angle as measured from the surface normal \cite{SSR_Greber}. 
The initial cloud diameter and the probing electron spot were left as free parameters.
The resulting electric field of the space charge cloud and its image along the probe electron trajectory is determined in summing the contributions of idividual cloud-electron trajectories, 
where the initial conditions were determined by a Monte-Carlo algorithm that satisfies the above described model parameters.
Integration of this field along the probe trajectory in turn delivers the kinetic energy gain of the probe electrons as a function of the delay $t_D=t_{probe}-t_{pump}$. 
Comparison of the simulation with the experiment identifies an effective number of electrons in the space charge cloud. 
For the data shown in \fig{F4} it is found to be $6.3\pm0.2\cdot10^4$ electrons in an initial disk with a diameter of $\unit{1.8}{\milli\meter}$ and a probe spot with a diameter smaller than $\unit{0.6}{\milli\meter}$.
In \fig{F4} we also show for comparison a simulation for a cloud with $\unit{400}{\micro\meter}$ initial diameter and a point like probe with 5000 electrons, which gives a limit of the shortest possible acceleration transients of $\unit{100}{\pico\second}$.
The resulting cloud currents from the model without acceleration were compatible with the measured emission currents, though are incompatible with the Richardson-Dushman formula (\eq{E1}).
In particular the diameter of the initial cloud appears too large, compared to the laser focus, which is in the order of 0.05Ê mm$^2$.
We take these deficiencies as a confirmation that the model assumption of no acceleration of the cloud is $\it{not}$ correct, i.e. as a clue for a Coulomb explosion, which accelerates the electrons in the cloud. 
It is beyond the scope of this paper to model the explosion in detail, also because more experiments with different probe electron energies should provide a larger data base for the test of the models. 
It is e.g. not clear whether the image charge evolves without delay.
Nevertheless, it has to be said that the average electron electrostatic energy in a homogeneously charged disk with a diameter of a laser focus of $\unit{0.25}{\milli\meter}$ diameter and  $6.3\cdot 10^4$ electrons  is $\unit{1.5}{\electronvolt}$, which is close to the measured average kinetic energy. 
Together with the electron temperatures after laser heating in the order of $\unit{3500}{\kelvin}$ the effective number of electrons  in the observed clouds is again compatible with the Richardson-Dushman formula.

%\section{CONCLUSIONS}

In conclusion it is shown that a time and energy resolved low energy electron scattering experiment  gives new and complementary insight into the dynamics of an expanding space charge cloud. 
The measured cloud energy and the transients in the order of $\unit{1}{\nano\second}$ indicate that a Coulomb explosion self-accelerates a space charge cloud which is generated by $\unit{}{\micro\joule}$ femtosecond laser pulses.
If this time resolved electron scattering experiment is expanded to its original idea, i.e. diffraction of low energy electrons (LEED) it will also become useful for the recording of structural changes on surfaces at ultra short time scales.

%\section{ACKNOWLEDGMENTS}
This project profited from skillful assistance of Jurt B\"{o}siger, Martin Kl\"{o}ckner and Hansruedi Scherrer. Finacial support from the Deutsche Forschungsgemeinschaft and the Swiss National Science Foundation, within the CERC3 network are greatfully acknowledged.

\setcounter{enumiv}{0}

\end{document}